\documentstyle[prl,aps,floats]{revtex}
\begin{document}
\twocolumn
\wideabs{
\title{The Harmonic Oscillator in the Plane and the Jordan - Schwinger
Algebras}

\author{Rabin Banerjee},
 
\address{S. N. Bose National Centre for Basic Sciences \\
JD Block, Sector III, Salt Lake City, Calcutta -700 098, India.
 }
\author{ Pradip Mukherjee}
 
\address{Presidency College\\
College Street, Calcutta -700 073, India.
}
\maketitle
\begin{abstract}
{The direct and indirect Lagrangian 
representations of the planar harmonic oscillator
have been discussed. The reduction of these Lagrangians in their basic
forms characterising either chiral, or 
pseudo - chiral oscillators have been given. A Hamiltonian analysis, showing
its equivalence with the Lagrangian formalism has also been provided.
Finally, we show that the chiral and pseudo - chiral modes act as
dynamical structures behind the Jordan - Schwinger realizations of the
SU(2) and SU(1,1) algebras. Also, the SU(1,1) construction found here
is different from the standard Jordan - Schwinger form.
 }\end{abstract}
}
The inverse problem of variational calculus is to construct the Lagrangian
from the equations of motion. Different Lagrangian representations are 
obtained from the direct and indirect approaches \cite{S}. In the direct
 representation
as many variables are introduced as there are in the equations of motion.
The equation of motion corresponding to a coordinate $q$ is related with the
variational derivative of the action with respect to the same coordinate.
Whereas, in the indirect representation, the equation of motion is supplemented
by its time - reversed image. The equation of motion with respect to the
original variable then corresponds to the variational derivative of the
action with respect to the  image coordinate and vice versa \cite{Ba}. The
linear harmonic oscillator, whose direct representation is completely
familiar, can also be solved 
in the inverse representation \cite{ta}. The importance
of the harmonic oscillator problem in physics hardly needs to be emphasised.
A thorough analysis of the different Lagrangian representations is thus
interesting in its own right. Note that due to the doubling of degrees of freedom associated with the indirect approach, the bidimensional oscillator appear
automatically. Thus a comparision of the different approaches can only been
done in the context of the two - dimensional (planar) oscillator.

   The planar harmonic oscillator model serves as a prototype in the
Landau problem of the motion
of a charged particle in an external magnetic field and thus is instrumental
in the theory of the Quantum Hall Effect \cite{G}. Now the usual 
Lagrangian of the harmonic oscillator in the plane can be viewed as a 
combination of the left and right handed 
chiral oscillators \cite{BG}. The later point has been 
exploited in \cite{RL} to discuss the noncommutativity of the Landau
problem. Various facets of the two - dimensional oscillator are then 
worth exploring from different angles.

       A beautiful piece of scenary that may be added here is the 
Jordan - Schwinger (J - S) construction of the SU(2) algebra from 
a pair of independent harmonic
oscillator algebras \cite{SK}. Such constructions can be generalised to
groups with polynomial algebras \cite{SBJ}. A physical interpretation of
 these algebraic constructions in terms of dynamical structures
is still lacking.

          In the present paper 
we consider the problem of the harmonic oscillator
in the 
 plane from the above mentioned points of views.
 The organisation of the paper is as follows. In section 1 we review 
the direct and indirect Lagrangian constructions of the linear harmonic
oscillator. Section 2 contains our analysis of the bidimensional oscillator
in the Lagrangian formalism. The reduction of the planar oscillator to their
basic forms characterising 
chiral and pseudo - chiral oscillators corresponding to
direct and indirect representations is discussed here.
 This is followed by a Hamiltonian analysis
in section 3 where an equivalence with the Lagrangian
formalism is also established.
 The connection of the J - S realizations of SU(2) and
 SU(1,1) algebras with
the elementary modes of the bidimensional oscillator - the
chiral and pseudo - chiral oscillators, respectively - 
 is discussed in section
4. Section 5 contains the concluding remarks.
{\vspace {.7cm}}

{\center{ \large{{\bf Section 1 : Direct and indirect Lagrangian representations of the harmonic
oscillator}}}
}

{\vspace {.7cm}}

The equation of motion satisfied by the harmonic oscillator is
\begin{equation}
\ddot{x} + \omega^2 x = 0\label{1}
\end{equation}
To find the Lagrangian in the direct method we write (\ref{1}) in the 
self adjoint form 
\begin{equation}
\frac{d}{dt}\dot{x} + \omega^2 x = 0\label{2}
\end{equation}
The variation of the action $S$ must then be
\begin{equation}
\delta S = \int^{t_2}_{t_1}dt\left[\frac{d}{dt}\dot{x} + \omega^2 x\right]
                          \delta x\label{3}
\end{equation}
so that the variational derivative of $S$ w.r.t. $x$ gives the
required equation of motion (\ref{1}). By discarding a surface term we get
from (\ref{3})
\begin{equation}
\delta S = -\delta\left(\int^{t_2}_{t_1}dt\left(\frac{1}{2}\dot{x}^2 -
    \frac{1}{2} \omega^2 x^2\right)\right)
                          \label{4}
\end{equation}
From (\ref{4}) we can easily identify the Lagrangian
\begin{equation}
L_D = \frac{1}{2}\dot{x}^2 -
    \frac{1}{2} \omega^2 x^2
                          \label{5}
\end{equation}
of the one dimensional harmonic oscillator.

  In the indirect approach we consider (\ref{1}) along with its 
time - reversed copy
\begin{equation}
\ddot{y} + \omega^2 y = 0\label{6}
\end{equation} 
and write the variation of the action as
\begin{equation}
\delta S = \int^{t_2}_{t_1}dt\left[\left(\frac{d}{dt}\dot{x} 
               + \omega^2 x\right)\delta y +
                \left(\frac{d}{dt}\dot{y} + \omega^2 y\right)\delta x\right]
                          \label{7}
\end{equation}
From (\ref{7}), equation (\ref{1}) is obtained by varying $S$ with $y$
whereas (\ref{6}) follows from varying $S$ with $x$.
Since the equations of motion for $x$ and $y$ follow as Euler - Lagrange 
equations for $y$ and $x$ respectively, the method is called the indirect
method. Now, starting from (\ref{7}) we can deduce
\begin{equation}
\delta S = -\delta \int^{t_2}_{t_1}dt\left[\dot{x}\dot{y} -\omega^2 x  y\right]
                          \label{8}
\end{equation}
It is then possible to identify
\begin{equation}
L_I = \dot{x}\dot{y} -\omega^2 x  y
                          \label{9}
\end{equation}
as the appropriate Lagrangian in the indirect representation.
The Lagrangian (\ref{9}) can be written in a suggestive form by the 
substitution of the hyperbolic coordinates $x_1$ and $x_2$
\cite{BGPV} defined by 
\begin{eqnarray}
x &=& \frac{1}{\sqrt{2}}(x_1 + x_2)\nonumber\\
y &=& \frac{1}{\sqrt{2}}(x_1 - x_2)\label{10} 
\end{eqnarray}
We find that the Lagrangian $L_{I}$ becomes
\begin{equation}
L_{I} = \frac{1}{2}\dot{x}_1^2 - \frac{\omega^2}{2}x_1^2
- \frac{1}{2}\dot{x}_2^2 + \frac{\omega^2}{2}x_2^2\label{111}
\end{equation}
The above Lagrangian can be expressed in a notationally elegant form \cite{BM}
\begin{equation}
L_I = \frac{1}{2}g_{ij}\dot{x}_i\dot{x}_j - \frac{\omega^2}{2}g_{ij}x_i x_j
                  \label{11}
\end{equation}
by introducing the pseudo - Eucledian metric $g_{ij}$ 
given by $g_{11} = -g_{22}
= 1$ and $g_{12}$ = 0.
{\vspace {0.7cm}}

{\center{\large{{\bf Section2 : Elementary modes of the bidimensional oscillator in direct 
and indirect representations}
}}}

{\vspace {.7cm}}

The Lagrangians given by equation (\ref{11}) represent the combination 
of two one dimensional oscillators, the equations of motion of which are
given by (\ref{1}) and (\ref{6}), in the indirect representation.
If we substitute $g_{ij}$ in (\ref{11}) by $\delta_{ij}$ then the Lagrangian
$L_I$ is mapped to
\begin{equation}
L_I \to  \frac{1}{2}\dot{x}_i\dot{x}_i - \frac{\omega^2}{2}x_i x_i
                  \label{12}
\end{equation}
Relabeling $x = x_1$ and $y = x_2$ and then using (\ref{5}) we
see that (\ref{12}) is the Lagrangian of the bidimensional oscillator
in the direct representation. Thus the mapping
\begin{equation}
g_{ij} \to \delta_{ij}\label{13}
\end{equation}
allows us to transform the indirect Lagrangian of the two - dimensional
oscillator in the hyperbolic coordinates to the direct Lagrangian in the
usual coordinates. The converse is also true. This helps us to find
the elementary modes in the indirect representation from the known result
for the direct Lagrangian (\ref{12}) \cite{BG}.

 The elementary modes of the bidimensional oscillator (\ref{12}) are \cite{BG}
\begin{equation}
L_{\pm} = \pm\omega\epsilon_{ij}x_i\dot{x}_j - \omega^2 x_i x_i\label{14}
\end{equation}
This can be demonstrated by the soldering
formalism 
which has found applications in various contexts -  duality symmetric
electromagnetic actions were constructed \cite{BW}; implications in
higher dimensional bosonization were discussed \cite{BK2}; the doublet
structure in topologically massive gauge theories was revealed \cite{BK1};
a host of phenomena in two dimensions were analysed \cite{A}.
We start from a simple sum
\begin{equation}
L(y,z) = L_+(y) + L_-(z)\label{s1}
\end{equation}
and consider the gauge transformation
\begin{equation}
\delta y_i= \delta z_i = \Lambda_i(t)\label{LL5}
\end{equation}
where $\Lambda_i$ are some arbitrary functions of time.
Under these transformations the change in $L$ is given by
\begin{eqnarray}
\delta L(y,z)& =&\delta L_+(y) + \delta L_-(z)\nonumber\\
             & =&\Lambda_i \left(J_{i}^{+}(y) + J_{i}^{-}(z)\right)\label{s2}
\end{eqnarray}
where the currents are,
\begin{equation}
J^{\pm}_i(x)
             = 2\left(\pm \omega\epsilon_{ij}\dot{x}_j - \omega^2x_i\right)
                      \label{CN1}
\end{equation}
The idea is to iteratively modify $L(y,z)$ by suitably introducing auxiliary
variables such that the new Lagrangian is invariant under the transformations
(\ref{LL5}). To this end
an auxiliary field $B_i$ transforming as (\ref{LL5}),\begin{equation}
\delta B_i = \Lambda_i\label{LL55}
\end{equation}
is introduced and a modified Lagrangian is constructed as
\begin{eqnarray}
L(y,z,B) &=& L(y,z) - B_i(J^{(+)}_i(y)+
              J^{(-)}_i(z))\nonumber\\
         &-& 2\omega^2 B_iB_i\label{LL8}
\end{eqnarray}
This Lagrangian is
 now invariant under (\ref{LL5}) and (\ref{LL55}).
Since the variable $B_i$ has no independent dynamics, it is eliminated by using
its equation of motion. The residual Lagrangian no longer depends on
$y$ or $z$ individually but only on the difference $y - z$. Writing this
difference as $x$, the residual Lagrangian reproduces (\ref{12}).
 
 The essence of the soldering procedure can be 
understood also in the following alternative way.
 Use $x_i = y_i - z_i$ in $L(y,z)$ to eliminate
$z_i$ so that
\begin{eqnarray}
L(y,x) = &-&2\omega\epsilon_{ij}y_i \dot{x}_j -\omega x_i \dot{x_j}
         \nonumber\\
        &-& 2\omega^2\left(y_i y_i - y_i x_i  + \frac{1}{2}x_i x_i\right)
        \label{sn5}
\end{eqnarray}
Since there is no kinetic term for $y_i$ it is really an auxiliary variable.
Eliminating $y_i$ from $L(y,x)$ by using its equation of motion
we directly arrive at (\ref{12}).
 Note that the opposite chirality of the elementary
Lagrangians are crucial in the cancellation of the time derivative of $y$
in (\ref{sn5}) which in turn is instrumental in the success of the soldering
method.
 
   The bidimensional Lagrangian (\ref{12}) is invariant under
\begin{equation}
x_i \to x_i + \theta\epsilon_{ij}x_j\label{20}
\end{equation}
which is nothing but an infinitesmal rotation in the $x_1,x_2$ plane.
The chiral oscillators (\ref{14}) are also separately invariant under
the SU(2) transformation (\ref{20}). 
Corresponding to this symmetry the angular momentum 
\begin{equation}
{\cal{J}} = \epsilon_{ij}x_ip_j\label{200}
\end{equation}
is conserved where
$p_i = -\omega\epsilon_{ij}x_j$ is the canonical momentum conjugate to
$x_i$.The spectrum of 
$\cal{J}$ is given by
\begin{equation}
{\cal{J}}_{\pm} =\pm \frac{\tilde{H}}{\omega}\label{22}
\end{equation}
where $\tilde{H}$ = $\omega^2(x_1^2 + x_2^2)$ is the Hamiltonian following from
$L_{\pm}$. The chiral oscillators thus manifest dual aspects of the
symmetry of (\ref{12}). This is why the 'plus' and 'minus' type of oscillators
 given by (\ref{14}) are interpreted as the left
and right handed chiral oscillators.

 Now, in view of the mapping (\ref{13}) existing between (\ref{11}) and 
(\ref{12}) one can construct the elementary modes of (\ref{11}) as has
been done for (\ref{12}).
By exactly a similar approach as detailed above one can show that
the elementary modes
\begin{equation}
{\cal{L_{\pm}}} = 
\pm i\omega\epsilon_{ij}x_i\dot{x}_j - \omega^2 g_{ij}x_i x_j\label{23}
\end{equation}
can be soldered to yield
 (\ref{11}). 
Of course the form (\ref{23}) is suggested by (\ref{14}), on account of
the mapping (\ref{13}). 
Note, however, that there is a factor of $i$ in the Lagrangians (\ref{23}) 
which makes them complex valued. Such Lagrangians
have recently appeared in a new canonical formulation of the damped
harmonic oscillator problem \cite{BM}.

Note that the composite Lagrangian (\ref{11}) and also the pieces (\ref{23})
are invariant under the transformation
\begin{equation}
x_i \to x_i + \theta\sigma_{ij}x_j\label{24}
\end{equation}
where $\sigma$ is the first Pauli matrix. The Noether charges following from
(\ref{23}) are
\begin{equation}
C_{\pm} = \pm\frac{\tilde{{\cal{H}}}}{\omega}\label{222}
\end{equation}
where $\tilde{{\cal{H}}}$ = $\omega^2(x_1^2 - x_2^2)$ is the 
Hamiltonian following from
${\cal{L_{\pm}}}$. The transformations (\ref{24}) belong to the SU(1,1) group.
  The doublets (\ref{23}) have opposite 'charges' w.r.t. the
SU(1,1) transformations in the plane. They may be aptly called as
 the pseudo - chiral oscillators.
The pseudo - chiral oscillators thus manifest dual aspects of the
SU(1,1) symmetry of (\ref{11}).
{\vspace {0.7cm}}

{\center{\large{{\bf Section 3 : Hamiltonian analysis}
}}}

{\vspace {.7cm}}

In the previous section we have seen that the Lagrangian of the bidimensional
oscillator in the usual coordinates is the synthesis of the chiral 
oscillators (\ref{14}) whereas the indirect Lagrangian in the hyperbolic
coordinate can be viewed as a 
 coupling of the independent pseudochiral doublet
(\ref{23}). This can also be understood from the Hamiltonian analysis.

  From the Lagrangian (\ref{12}) we can construct the
Hamiltonian by a Legendre transformation
\begin{equation}
H_D = \frac{1}{2} p_ip_i + \frac{\omega^2}{2}x_ix_i\label{26}
\end{equation}
where $p_i$ are canonically conjugate to $x_i$. $H_D$ is manifestly the
sum of two one - dimensional Harmonic oscillators,
\begin{equation}
H_D = H_1 + H_2\label{b}
\end{equation}
where
\begin{equation}
H_i = \frac{1}{2} p_i^2 + \frac{\omega^2}{2}x_i^2;\hspace{.2cm}(i = 1,2)
     \label{261}
\end{equation}
We will show that the individual pieces $H_{i}$ are the Hamiltonians of
 the left and right chiral oscillators.
Consider the Lagrangian $L_+$ of the left chiral oscillator
(the first one of (\ref{14})). This is already in the first order form
and we can read off the Hamiltonian directly
\begin{equation}
H_+ = \omega^2 \left(x_1^2 + x_2^2 \right)\label{hn1}
\end{equation}
with the symplectic algebra
\begin{equation}
\{x_i, x_j\} = -\frac{1}{2\omega}\epsilon_{ij}\label{bn1}
\end {equation}
From (\ref{bn1}) we find that $2\omega x_1$ is canonically conjugate to $x_2$.
Now by a canonical transformation to the set $(x,p_x)$ defined by
\begin{equation}
x_1 = \frac{1}{\sqrt{2}}\frac{p_x}{\omega}\hspace{.3cm};\hspace{.3cm}
x_2 = \frac{1}{\sqrt{2}}x\label{tn2}
\end{equation}
the Hamiltonian (\ref{hn1}) becomes
\begin{equation}
{H}_+ = \left(\frac{p_x^2}{2} + \frac{\omega x^2}{2} \right)\label{hn2}
\end{equation}
The above Hamiltonian coincides with $H_1$ of (\ref{b}). Similarly we can
prove that the piece $H_2$ of (\ref{b}) follows from $L_-$ of (\ref{14}).
The reduction of the bidimensional oscillator
in the direct representation
as a doublet of the chiral oscillators is thus also established 
in the Hamiltonian approach.

 We then consider the indirect representation (\ref{11}). The  Hamiltonian
obtained from (\ref{11}) is explicitly
\begin{equation}
H_I = \left(\frac{1}{2}p_1^2 +\frac{1}{2}\omega^2x_1^2\right) 
- \left(\frac{1}{2}p_2^2 +\frac{1}{2}\omega^2x_2^2\right)\label{27}
\end{equation}
Note that it is equivalent to the {\it{difference}}
 of two one - dimensional oscillators. 
Making a canonical transformation
\begin{eqnarray}
p_{\pm} =\frac{1}{\sqrt{2}}p_1 \pm i
             \frac{\omega}{\sqrt{2}}x_2\nonumber\\
x_{\pm} =\frac{1}{\sqrt{2}}x_1 \pm i
             \frac{1}{\sqrt{2}}\frac{p_2}{\omega}\label{ct}
\end{eqnarray} 
it is possible to write
\begin{equation}
H_I = {\cal{H_+}} + {\cal{H_-}}\label{29}
\end{equation}
where
\begin{equation}
{\cal{H_{\pm}}} = \frac{1}{2}p_{\pm}^2 +\frac{1}{2}\omega^2 x_{\pm}^2\label{30}
\end{equation}
The price one has to pay is that the canonical variables $x_{\pm}$
and $p_{\pm}$ are no longer real. As a result
 the Hamiltonians $ {\cal{H_{\pm}}}$ are not hermitian. Note, however, that
\begin{equation}
{\cal{H_{\pm}}}^{\dagger}= {\cal{H_{\mp}}}\label{31}
\end{equation}
so that the hermiticity of $H_I$ is preserved.
One can prove that
\begin{equation}
\eta {\cal{H_{\pm}}} \eta^{-1} = {\cal{H_{\pm}^{\dagger}}}\label{ph}
\end {equation}
where $\eta = PT$. The above condition follows from the
observation
that under $PT$ transformation
\begin{equation}
\eta x_i \eta^{-1} = g_{ij}x_j,\hspace{.3cm} \eta p_i \eta^{-1} =  -g_{ij}p_j
\end{equation}
Hence  
\begin{equation}
\eta x_{\pm} \eta^{-1} = x^{\dagger}\hspace{.3cm}{\rm{and}}\hspace{.3cm}
   \eta p_{\pm} \eta^{-1} =  -p^{\dagger}\label{pc}
\end{equation}
Basing on (\ref{ph}) it is possible to
build a consistent quantum mechanics \cite{BM}.
 Conservation of probability is ensured by the following
redefinition of the scalar product
\begin{equation}
<<\psi|\phi>> = \int d\tau \tilde{\psi}\phi\label{311}
\end{equation}
where $\tilde{\psi}$ is the $\eta$ - transformed wavefunction.
Operators satisfying (\ref{ph}) are called generalised hermitian (g - hermitian)
operators \cite{D}. 
One can define the g - hermitian adjoint of an operator O
by
\begin{equation}
\tilde{O} = \eta^{-1} O^{\dagger} \eta\label{phc}
\end{equation}
Then the idea of g - hermiticity can easily be seen as a generalisation
of the usual idea of hermiticity.
 Recently this
idea has resurfaced again\cite{BAM} where the term pseudo hermiticity
is used to denote g - hermiticity.

We can now show that the Hamiltonians ${\cal{H_{\pm}}}$
follow from the Lagrangians (\ref{23}) in the same way as 
we have shown that the Hamiltonians $H_1$ and $H_2$ of equation (\ref{b})
 follow from the
Lagrangians of the chiral oscillators (\ref{14}).
Hence we show by hamiltonian analysis that in the indirect representation 
 the bidimensional oscillator is reduced to a combination of two
 pseudo - chiral oscillators.
{\vspace {.7cm}}

{\center{\large{{\bf Section4 : The chiral and the pseudo - chiral oscillators as the 
dynamical structures behind the Jordan - Schwinger realizations of SU(2) and 
SU(1,1) algebras}
}}}

{\vspace {.7cm}}

  We begin with the Jordan - Schwinger (J - S) realization of the SU(2) algebra.
Define two sets of operators $(a,a^{\dagger})$ and $(b,b^{\dagger})$ which
satisfy the bosonic algebra
\begin{equation}
[a,a^{\dagger}] = [b,b^{\dagger}] = 1\label{33}
\end{equation}
The algebras are independent;
\begin{equation}
[a,b^{\dagger}] = [a,b] = 0\label{331}
\end{equation}
Using the a and b set of operators one can construct
\begin{eqnarray}
J_z &=& \frac{1}{2}\left(a^{\dagger}a - b^{\dagger}b\right)\nonumber\\
J_+ &=& J_x + iJ_y = a^{\dagger}b \nonumber\\
J_- &=& J_x - iJ_y = a b^{\dagger}\label{34}
\end{eqnarray}
The operators defined by (\ref{34}) satisfy the SU(2) algebra
\begin{equation}
\left[J_z,J_{\pm}\right] = \pm J_{\pm},\hspace{.3cm}\left[J_+,J_-\right]=2J_z
                    \label{341}
\end{equation}
These operators constitute the J - S realization of SU(2).

 The J - S realization of the SU(2) algebra is constructed from two 
independent oscillator algebras. 
It is now pretty suggestive
 that the chiral oscillators (\ref{14}) constitute
a dynamical model for the construction (\ref{34}). Indeed, from (\ref{22})
we observe that corresponding to one quantum of excitation of the left
handed chiral oscillator (energy $\omega$) 
the angular momentum projection is $\frac{1}{2}$
whereas that for the right handed one it is $-\frac{1}{2}$. A physical system
corresponding to these excitations is also there, namely an electron rotating
clockwise or anticlockwise
about the direction of the external magnetic field. The operators of
(\ref{34}) can be explicitly constructed by the angular momentum addition
rules. Using (\ref{22}) and the expression of the harmonic oscillator
Hamiltonian in terms of the creation and annihilation operators we
can write
\begin{eqnarray}
J_{az} &=& \frac{1}{2}a^{\dagger}a\nonumber\\
J_{bz} &=& -\frac{1}{2}b^{\dagger}b \label{342}
\end{eqnarray}
Note that we have deducted the zero - point energy.
The $ z$ component of the total angular momentum is
\begin{equation}
J_z = J_{az} + J_{bz}\label{3421}
\end{equation}
which is the $J_z$ given by equation (\ref{34}).
 Again the total angular
momentum operator squared is given by
\begin{eqnarray}
J^2& =& j(j + 1)\nonumber\\
   & =&\left[\frac{1}{2}\left(a^{\dagger} a + b^{\dagger}b\right)\right]
            \left[\frac{1}{2}\left(a^{\dagger}a + b^{\dagger}b\right)
                        + 1\right]\label{343}
\end{eqnarray}
Now using the relation
\begin{eqnarray}
J^2& =& J_x^2 + J_y^2 + J_z^2\nonumber\\
   & =& J_{z}^2 + \frac{1}{2}\left(J_+J_- + J_-J_+\right)\label{344}
\end{eqnarray}
and the expression for $J_z$
 we can deduce the expressions of $J_{\pm}$ given in (\ref{34}).
The identification of (\ref{14}) as the dynamical structures behind the
J - S algebraic construction is thus fully established.

  The chiral oscillators carry dual aspects of SU(2) and provide 
dynamical structures for the J - S consruction of SU(2). Clearly, the
pseudo chiral oscillators (\ref{23}) are candidates for SU(1,1) as
they also serve as the 'spin' doublets with respect to SU(1,1) transformations.
We introduce
\begin{equation}
a = \sqrt{\frac{\omega}{2}}\left(x_+ + \frac{ip_+}{\omega}\right)\label{37}
\end{equation}
and
\begin{equation}
b = \sqrt{\frac{\omega}{2}}\left(x_- + \frac{ip_-}{\omega}\right)\label{38}
\end{equation}
The g - hermitian conjugates of $a$ and $b$ are obtained from the
definition (\ref{phc}) as 
$\tilde{a}$ and $\tilde{b}$ respectively where 
\begin{equation}
\tilde{a} = \eta ^{-1} a^{\dagger} \eta\label{pha}
\end{equation}
\begin{equation}
\tilde{b} = \eta ^{-1} b^{\dagger} \eta\label{pha1}
\end{equation}
From (\ref{37}) and (\ref{38}) we get using (\ref{pc})
\begin{equation}
\tilde{a} = \sqrt{\frac{\omega}{2}}\left(x_+ - \frac{ip_+}{\omega}\right)
                                 \label{371}
\end{equation}
and
\begin{equation}
\tilde{b} = \sqrt{\frac{\omega}{2}}\left(x_- - \frac{ip_-}{\omega}\right)
                                  \label{381}
\end{equation}
It is easy to prove the algebra
\begin{equation}
[a,\tilde{a}] = [b,\tilde{b}] = 1\label{33n}
\end{equation}
and
\begin{equation}
[a,\tilde{b}] = [a,b] = 0\label{33n1}
\end{equation}
remembering that ($x_+,p_+$) and ($x_-,p_-$) are independent canonical pairs.
Let us then construct
\begin{eqnarray}
J_z &=& \frac{1}{2}\left(\tilde{a}a - \tilde{b}b\right)\nonumber\\
J_+ &=& \tilde{a} b\nonumber\\
J_- &=& -\tilde{b}a \label{40}
\end{eqnarray}
where,
\begin{equation}
J_{\pm} = J_x \pm i J_y
\label{pm}
\end{equation}
One can easily verify that these operators satisfy
\begin{equation}
\left[J_z,J_{\pm}\right] = \pm J_{\pm},\hspace{.3cm}\left[J_+,J_-\right]=-2J_z
                    \label{41}
\end{equation}
which is nothing but the SU(1,1) algebra. The construction (\ref{40})
is then a realization of the SU(1,1) algebra based on the algebra (\ref{33n})
and (\ref{33n1}). The operators $a$, $\tilde{a}$($b$, $\tilde{b}$) are
respectively the annihilation and creation operators belonging to the
plus (minus) type pseudo - chiral oscillators (\ref{23}). The corresponding 
Hamiltonians has been shown to be 
${\cal{H_{\pm}}}$ given by 
(\ref{30}). These Hamiltonians
can be diagonalized in terms of the operators $a$ and $b$ as 
\begin{equation}
{\cal{H_+}} = \omega \tilde{a} a
\end{equation}
and
\begin{equation}
{\cal{H_-}} = \omega \tilde{b} b
\end{equation}
where, again, we have subtracted the zero point energy.
Now, in view of the relation (\ref{222}), the construction (\ref{40}) can be 
interpreted in terms of the pseudo - chiral oscillators just as (\ref{34})
was interpreted in terms of the chiral oscillators. Note, however,
 a difference of 
minus sign in the last equation of 
(\ref{40}) and (\ref{34}). This is due to the specific form
of the Casimir operator of SU(1,1)
\begin{eqnarray}
J^2 &=& J_z^2 - \frac{1}{2}( J_+J_- + J_-J_+)\nonumber\\
    &=& \left[\frac{1}{2}\left(\tilde{a}a + \tilde{b}b\right)\right]
     \left[\frac{1}{2}\left(\tilde{a}a + \tilde{b}b\right) + 1 \right]
       \label{c2}
\end{eqnarray}
Comparing the above with the well known form of the
Casimir operator of SU(2) ( see equation (\ref{344}))
, we can understand the difference, mentioned above,
 between (\ref{40}) and (\ref{34}).
The structural similarity between the expressions of the Casimir operator
in terms of the basic variables 
in (52) and (\ref{c2}) is remarkable. This reveals again the parallel
between our constructions of SU(2) and SU(1,1) algebras based on the
dynamical structures of the chiral or pseudo - chiral oscillators.
Using (\ref{pm}) we can explicitly determine $J_x$, $J_y$ and $J_z$
from (\ref{40}) as
\begin{eqnarray}
J_x &=& \frac{1}{2}\left(\tilde{a}b - \tilde{b}a\right)\nonumber\\
J_y &=& -\frac{i}{2}\left(\tilde{a}b + \tilde{b}a\right)\nonumber\\
J_z &=& \frac{1}{2}\left(\tilde{a}a - \tilde{b}b\right)\label{x}
\end{eqnarray}
Using the expressions (\ref{37}), (\ref{38}),(\ref{371}) and (\ref{381})
we find that $J_x$ is hermitian whereas $J_y$ and $J_z$ are anti -  hermitian.
It is possible to construct a realization of the SU(1,1) algebra 
consisting of hermitian operators only from (\ref{x}) by the following 
mapping
\begin{eqnarray}
J_x &\to & J_y\nonumber\\
J_y &\to & iJ_z\nonumber\\
J_z &\to & -iJ_x\label{map2}
\end{eqnarray}
Explicitly,
\begin{eqnarray}
J_x &=& \frac{i}{2}\left(\tilde{a}a - \tilde{b}b\right)\nonumber\\
J_y &=& \frac{1}{2}\left(\tilde{a}b - \tilde{b}a\right)\nonumber\\
J_z &=& -\frac{1}{2}\left(\tilde{a}b + \tilde{b}a\right)\label{x1}
\end{eqnarray}
 That the mapping (\ref{map2}) preserves the SU(1,1)
algebra can be seen from (\ref{41}). Alternatively, one can check it
directly from (\ref{x1}).

   At this point it is instructive to compare our representation (\ref{40})
with the usual J - S realization of SU(1,1). The later is given by \cite{SBJ}
\begin{eqnarray}
J_z &=& \frac{1}{2}\left(a^{\dagger}a + bb^{\dagger}\right)\nonumber\\
J_+ &=& J_x + iJ_y = a^{\dagger}b^{\dagger} \nonumber\\
J_- &=& J_x - iJ_y = a b\label{340}
\end{eqnarray}
This representation is based on the algebras given by (\ref{33}) 
and (\ref{331}),
which are two independent harmonic oscillator algebras.
Note that, in contrast to (\ref{40}), (\ref{340}) 
cannot be interpreted in terms of independent
dynamical structures. This can be seen very simply by writing the Casimir
operator
from (\ref{340})
\begin{eqnarray}
C &=& J_z^2 - \frac{1}{2}\left(J_+J_- + J_-J_+\right)\nonumber\\
  &=& \frac{1}{4}\left(a^{\dagger}a - b^{\dagger}b\right)^2 - 
          \frac{1}{2}\left(a^{\dagger}a + b^{\dagger}b + 1\right)\label{c3}
\end{eqnarray}
Clearly this cannot be factorised as (52). On the otherhand, the 
Casimir operator
 obtained from our realization 
(\ref{40}) factorises properly.
 The realization (\ref{40}) is
therefore fundamentally different from the usual one ( equation (\ref{340}). 
{\vspace {.7cm}}

{\center{\large{{\bf Section 5 : Conclusions}
}}}

{\vspace {0.7cm}}
  The one dimensional harmonic oscillator is known to provide a fundamental
basis for solving a great variety of problems. Here we have analysed the
ramifications leading from the oscillator in the plane, arguably whose
best known appearence is in the context of the planar motion of an electron
moving under the influence of a constant perpendicular magnetic field
(Landau problem). The fundamental modes of the 2 - d oscillator were studied,
both in the direct and indirect representations. These modes were termed as the
chiral and pseudo - chiral oscillators, respectively. An equivalence between the Lagrangian and the Hamiltonian formulations was established. Finally
 it was shown that the basic operators defining the chiral (pseudo - chiral)
operators occured in the Jordan - Schwinger construction of SU(2) (SU(1,1))
algebras. This provided a dynamical realization of the algebras.
{\vspace {.7cm}}

{\large{{\bf Acknowledgement}}}
 
 One of the authors (P.M.) would like to thank Prof.
S. Dattagupta, Director, S. N. Bose National Center for Basic Sciences
for allowing him to work
as Visiting Associate (U. G. C.) in the Center. He also wishes to thank
the U. G. C. for the award of Visiting Associateship.

\end{document}